\documentclass[12pt]{article}   
\usepackage{epsfig}

\newlength{\dinwidth} \newlength{\dinmargin}
\begin{document}
\begin {flushright}
Cavendish-HEP-04/17
\end {flushright} 
\vspace{3mm}

\begin{center}
{\Large \bf Charged Higgs production with a top quark \\
at the LHC\footnote{Presented
at the DIS 2004 Workshop, Strbske Pleso, Slovakia, 14-18 April, 2004.}}
\end{center}
\vspace{2mm}

\begin{center}
{\large Nikolaos Kidonakis}\\
\vspace{2mm}
{\it Cavendish Laboratory, University of Cambridge\\
Madingley Road, Cambridge CB3 0HE, UK\\
E-mail: kidonaki@hep.phy.cam.ac.uk}
\end{center}

\vspace{3mm}

\begin{abstract}

I discuss charged Higgs production via the process
$bg \rightarrow t H^-$ at the LHC. I show that the cross section
is dominated by soft-gluon corrections and I provide results
for its dependence on the charged Higgs mass   
and on the scale, including higher-order effects.

\end{abstract}

\thispagestyle{empty} \newpage \setcounter{page}{2}

\section{Introduction} 

A future discovery of a charged Higgs boson would be a sure sign of 
new physics beyond the Standard Model. The Minimal Supersymmetric
Standard Model (MSSM) includes in its particle content charged Higgs bosons
in addition to neutral Higgs.
The discovery of the Higgs boson, including the charged Higgs of the MSSM,
is a major goal at the Tevatron and the LHC.

The LHC has good potential for discovery of a charged Higgs.
A promising channel is $bg \rightarrow tH^-$  
\cite{eH,lH,LesHouchesHiggs}.
The complete next-to-leading order QCD corrections to this process
have been recently derived in Refs. \cite{Zhu,Plehn}.

The Born cross section is proportional to $\alpha \alpha_s
(m_b^2\tan^2 \beta+m_t^2 \cot^2 \beta)$.
Here $\tan \beta=v_2/v_1$ is the ratio of the vacuum
expectation values of the two Higgs doublets in the MSSM.
We use the $\overline{\rm MS}$ running top and bottom quark masses, 
corresponding to pole masses $m_t=175$ GeV and $m_b=4.8$ GeV, in the
$\tan^2 \beta$ and $\cot^2 \beta$ terms, but set $m_b=0$ elsewhere.

\section{Charged Higgs NNLO-NLL cross section}

The production cross section for 
$b(p_b) + g(p_g) \longrightarrow t(p_t)+H^-(p_{H^-})$ at the LHC 
for large $m_{H^-}$ is actually dominated by soft-gluon 
corrections from the near-threshold region \cite{LesHouchesHiggs,NKrev}.
We define the standard kinematical invariants $s=(p_b+p_g)^2$, 
$t=(p_b-p_t)^2$, $u=(p_g-p_t)^2$, and $s_4=s+t+u-m_t^2-{m_{H^-}}^2$.
At threshold $s_4 \rightarrow 0$.
The threshold soft-gluon corrections then take the form
$[\ln^l(s_4/{m^2_{H^-}})/s_4]_+$, with $l \le 2n-1$ for  
the order $\alpha_s^n$ corrections.
They are calculated following the methods in 
Refs. \cite{NK1,NKuni}, which have been
applied to various processes \cite{NK2,LesHouchesQCD}. 
The leading logarithms (LL) are with $l=2n-1$ while
the next-to-leading logarithms (NLL) are with $l=2n-2$.
Here we calculate NLO and NNLO soft-gluon corrections corrections 
at NLL accuracy. We denote them as NLO-NLL and NNLO-NLL, respectively.
Thus, at NLO we include $[\ln(s_4/{m^2_{H^-}})/s_4]_+$ (LL) 
and $[1/s_4]_+$ (NLL) terms. Although we do not include the full virtual 
$\delta(s_4)$ terms, we include those $\delta(s_4)$ terms that involve the 
factorization and renormalization scales, denoted by $\mu$.
At NNLO, we include $[\ln^3(s_4/{m^2_{H^-}})/s_4]_+$ (LL)
and $[\ln^2(s_4/{m^2_{H^-}})/s_4]_+$ (NLL) terms. We also include 
some $[\ln(s_4/{m^2_{H^-}})/s_4]_+$ and $[1/s_4]_+$ terms
that involve the scale $\mu$ and some $\zeta_2$ and $\zeta_3$ constants.
Explicit analytical expressions are given in 
\cite{LesHouchesHiggs}.

\begin{figure}[!thb]
\vspace*{7.0cm}
\begin{center}
\includegraphics{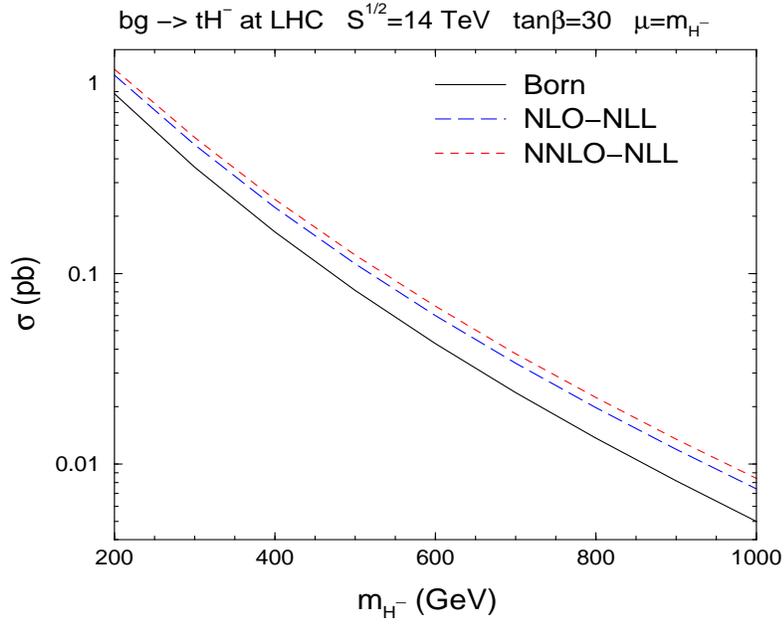}
\vspace{1.3cm}
\caption[*]{Charged Higgs production cross section versus charged Higgs mass.}
\label{fig1}
\end{center}
\end{figure}

In Fig. \ref{fig1} we plot the cross section for charged Higgs production 
via $bg \rightarrow t H^-$ at the LHC as a function of the charged Higgs mass
for $\tan \beta=30$ and $\mu=m_{H^-}$.
The approximate NNLO parton distributions of Ref. \cite{mrst2002} have 
been used.
We note that in the corresponding figures in 
Refs. \cite{LesHouchesHiggs,NKrev} 
the pole mass of the bottom quark was used; the $\overline{\rm MS}$ mass,
however, is a preferable choice \cite{Plehn}.  
The Born, NLO-NLL, and NNLO-NLL cross sections are shown.
The cross sections span over two orders of magnitude in the mass range
200 GeV $\le m_{H^-} \le$ 1000 GeV.

\begin{figure}[!thb]
\vspace*{7.0cm}
\begin{center}
\includegraphics{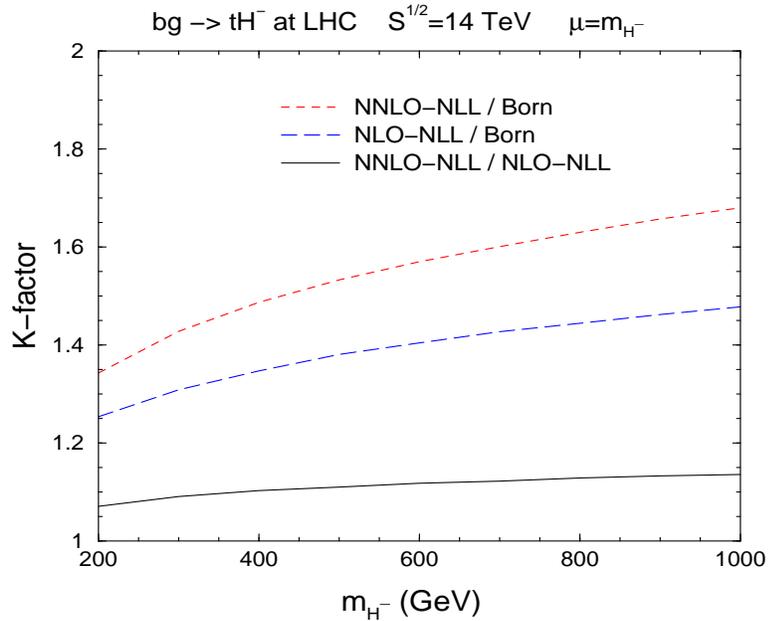}
\vspace{1.3cm}
\caption[*]{$K$-factors versus charged Higgs mass.}
\label{fig2}
\end{center}
\end{figure}

In Fig. \ref{fig2} we plot the $K$-factors, i.e. the ratios of 
the NLO-NLL and NNLO-NLL cross sections to the Born cross section,
and the ratio of the NNLO-NLL to the NLO-NLL cross section.
We use the same NNLO parton densities and couplings at all orders,
so as to concentrate on the effect of the soft-gluon corrections.
It should be stressed that the NLO-NLL/Born ratio is quite close to the 
exact NLO/Born ratio \cite{Zhu,Plehn} (note that different conventions
and scales are used there) which indicates that threshold
corrections indeed dominate the cross section. The difference between
the exact NLO and NLO-NLL results is only a few percent.
The NNLO-NLL corrections provide a significant enhancement
to the NLO cross section.

\begin{figure}[!thb]
\vspace*{7.0cm}
\begin{center}
\includegraphics{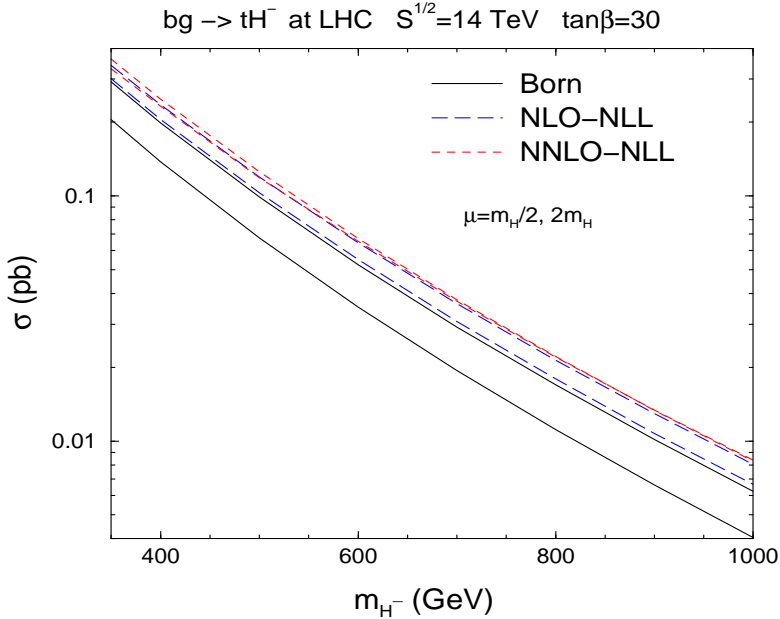}
\vspace{1.3cm}
\caption[*]{The scale dependence of charged Higgs production cross sections.}
\label{fig3}
\end{center}
\end{figure}

In Fig. \ref{fig3} we plot the cross section with two different 
choices of scale, $\mu=m_{H^-}/2$ and $2m_{H^-}$. 
For clarity we concentrate on $m_{H^-} \ge 350$ GeV.
We see that the variation 
with scale of the Born cross section is large. The variation at 
NLO-NLL is smaller, and at NNLO-NLL it is very small. 
In fact the two NNLO-NLL curves are on top of each other for most
of the range in $m_{H^-}$.
Hence, the scale dependence of the cross section is drastically reduced when
higher-order corrections are included.

Finally, we note that the cross section for 
${\bar b} g \rightarrow {\bar t} H^+$ is the same as for
$bg \rightarrow t H^-$. 

\section*{Acknowledgements} 

The author's research has been 
supported by a Marie Curie Fellowship of the European Community programme 
``Improving Human Research Potential'' under contract no.
HPMF-CT-2001-01221.

\end{document}